# Alterations in cell surface area and deformability of individual human red blood cells in stored blood

HyunJoo Park[1+], Misuk Ji[2+§], SangYun Lee, Kyoohyun Kim, Yong-Hak Sohn[2,*], Seongsoo Jang[3,*] and YongKeun Park[1,*]

[1]Department of Physics, Korea Advanced Institute of Science and Technology, Daejeon 305-701, Republic of Korea.
[2]Department of Laboratory Medicine, Eulji University Hospital, Daejeon 302–799, Republic of Korea.
[3]Department of Laboratory Medicine, University of Ulsan, College of Medicine and Asan Medical Center, Seoul 138–736, Republic of Korea.
[§]Current affiliation: Department of Laboratory Medicine, Konkuk University School of Medicine, Seoul 380-701, Republic of Korea.
[*]Corresponding Authors: medsohn@eulji.ac.kr, ssjang@amc.seoul.kr and yk.park@kaist.ac.kr

Manuscript information:
Text: 18 pages
Figures: 6
Supporting Information

*current aff

**Abstract**: The functionality and viability of stored human red blood cells (RBCs) is an important clinical issue in transfusion. To systematically investigate changes in stored whole blood, the hematological properties of individual RBCs were quantified in blood samples stored for various periods with and without a preservation solution called CPDA-1. With 3-D quantitative phase imaging techniques, the optical measurements of the 3-D refractive index (RI) distributions and membrane fluctuations were done at the individual cell level. From the optical measurements, the morphological (volume, surface area and sphericity), biochemical (hemoglobin content and concentration), and mechanical parameters (dynamic membrane fluctuation) were simultaneously quantified to investigate the functionalities and their progressive alterations in stored RBCs. Our results show that the stored RBCs without CPDA-1 had a dramatic morphological transformation from discocytes to spherocytes within 2 weeks which was accompanied with significant decreases in cell deformability and cell surface area, and increases in sphericity. However, the stored RBCs with CPDA-1 maintained their morphology and deformability for up to 6 weeks.

**Introduction**
Blood storage is important for emergency transfusions during surgery and for the treatment of anemia. According to current protocols, human red blood cells (RBCs) can be stored up to 42 days at 1–6°C in a preservation solution (49). This solution, containing citrate, phosphate, dextrose, and adenosine (CPDA-1), prevents blood coagulation by chelating calcium and supplies the nutrient glucose and adenosine triphosphate (ATP) while maintaining a low pH. However, the results of clinical trials have shown progressive deteriorations in the functionalities of blood during blood banking/storage, and strong correlations between the stored period of blood and the mortality of transfused patients (32, 46, 53). In clinical practice, it is generally recommended to use RBCs stored less than 14 days due to increase the risk of death and complications, although the shelf life of stored RBCs is much longer (26).

The reasons for the increased mortality associated with blood transfusions and longer storage of RBCs are not well understood (28). However, previous studies have reported adverse alternations in the properties of RBCs over several weeks of storage; the hematological test showed noticeable increases in RDW (red cell distribution width) while other parameters including the mean corpuscular volume, mean corpuscular hemoglobin and mean corpuscular hemoglobin concentration were not changed significantly (1). In addition, the rate of morphological changes from discocyte to spherocyte gradually increases and becomes 50% at 2 weeks (6). The relative biochemical parameters including 2,3-DPG, potassium, pH, and ATP and nitric oxide levels are also changed toward unfavorable conditions for RBC viability (5). Elongation index tests (6, 57), optical tweezers techniques (14, 18), and ektacytometry experiments (15, 16, 54) have revealed that stored RBCs lose deformability as a function of the storage period. Recent studies have reported a decrease in the dynamic membrane fluctuations of stored RBCs (8) and a delay in passing through a narrow microfluidic channel as a result of large cell deformations (57) over time in storage.

One of the important questions about stored RBCs is how these morphological and biochemical alternations from long storage periods affect the functions of RBCs. To address this question, systematic characterizations of individual RBCs should be quantified as a function of the storage time and at the individual cell level. Although previous studies on stored RBCs have significantly enhanced our understanding of storage lesions, none of them were able to investigate the morphological, biochemical, and mechanical parameters of individual RBCs simultaneously. Moreover, many of the previous methods only measured RBCs in the stored blood with CDPA-1 and thus not well suited to investigate the effects of the preservation solution on the properties of individual RBCs.

Here, we systematically investigated the alterations of biophysical properties of individual RBCs as a function of the storage duration. With 3-D quantitative phase imaging (QPI) techniques, 3-D refractive index (RI) tomograms and dynamic membrane fluctuations of individual RBCs were simultaneously and quantitatively measured based on the principles of laser interferometry and optical diffraction tomography (29, 41). From the optical measurements, the morphological (volume, surface area and sphericity), biochemical (hemoglobin content and concentration), and mechanical parameters (dynamic membrane fluctuation) of individual RBCs were simultaneously quantified to investigate the functionalities and their progressive alterations in stored RBCs. The measurements were performed every few days up to 6 weeks to systematically investigate the alterations of these RBC parameters over time in storage. To investigate the effects of the preservation solution on the RBC parameters, we tracked two groups of RBCs: one stored in the absence and the other in the presence of a preservation solution known as CPDA-1. Our results show that the rate of morphological transformation from discocytes to spherocytes increased as the storage time increased, which was accompanied with increased membrane stiffness. For the stored RBCs without CPDA-1, all the RBCs became spherical shape within a week, whereas the storage agent CPDA-1 prevented the morphological transformation of the RBCs and maintained a healthy level of dynamic membrane fluctuations in the RBCs for up to 6 weeks in storage.

**Results**
**Morphological alterations in RBCs over time in storage**
To study how the morphologies of the stored RBCs changes over time in storage in the absence and presence of the preservation solution CPDA-1, the 3-D RI tomograms of individual stored RBCs were measured with cDOT at days 1, 5, 13, 20, 27, 34 and 41 of the storage period. The RBCs were stored at 4°C. Representative RI tomograms are shown in Fig. 1. The measured 3-D RI tomograms of RBCs without CPDA-1 showed significant morphological changes over time in storage (Fig. 1A&B). The RBCs underwent morphological transitions from discocytes to discocytes by day 5 of storage. Bumpy and specular structures emerged in the RBC membranes, and their dimple structures in the center disappeared. After 2 weeks, all the RBCs without CPDA-1 became spherocytes, that is, the RBCs had spherical shapes.

In contrast, characteristic thin, doughnut-like shapes in the discocytes were still found in the RBCs stored with CPDA-1 up to 3 weeks of storage (Fig. 1C&D). After 20 days of storage, a portion of the spherocytes increased among the RBCs stored with CPDA-1. These morphological changes for both RBC groups over long storage periods were consistent with a previous report in which scanning electron microscopy was used (6). These morphological transitions can also be clearly seen in the rendered isosurface images of the reconstructed RI tomograms (Fig. 1 B&D).

For a quantitative analysis, the morphological parameters including the cell volume, cell surface area, and sphericity were calculated from the measured 3-D RI tomograms (see *Materials and Methods*). The results are shown in Fig. 2. During 41 days of the storage period, the volumes of RBCs were maintained for both the control and the RBCs with CDPA-1 (Fig. 2A&B). The mean values for the cellular volumes of the stored RBCs without CPDA-1 at days 1, 5, 13, 20, 27, 34, and 41 of the storage period were 89.7 ± 11.9 (n = 37), 87.0 ± 15.6 (n = 39), 80.7 ± 7.4 (n = 39), 83.2 ± 9.8 (n = 36), 82.8 ± 10.6 (n = 39), 89.9 ± 11.6 (n = 32), and 88.4 ± 9.9 fL (n = 34), and with CPDA-1 88.4 ± 12.2 (n = 43), 87.2 ± 13.9 (n = 44), 85.9± 11.0 (n = 47), 83.3 ± 11.9 (n = 45), 86.0 ± 12.0 (n = 37), 83.2 ± 12.6 (n = 43), and 82.5 ± 11.1 fL (n = 49), respectively. The unchanged cellular volume in the RBCs stored with CPDA-1 is consistent with a previous study using a complete blood machine, in which a large number of RBCs were averaged (1).

The surface areas of the RBCs stored without CPDA-1 started to decrease around day 5, and almost all RBCs started to have similar small sizes after 13 days of storage (Fig. 2C). In contrast, the surface areas of the RBCs stored with CPDA-1 started to gradually decrease over a long time exhibiting large cell-to-cell variations. The mean values for the surface areas of RBCs without CPDA-1 at the corresponding storage periods were 152.7 ± 14.7, 127.7 ± 27.8, 98.5 ± 6.6, 100.2 ± 8.5, 99.5 ± 9.0, 101.6 ± 9.1, and 101.0 ± 8.0 μm$^2$, respectively. At day 5, the surface areas decreased by 16 % and further decreased by 33% at day 13. The mean values for the surface areas of the RBCs with CPDA-1 at the corresponding storage periods were 152.4 ± 13.3, 142.8 ± 20.1, 112.0 ± 23.5, 111.7 ± 17.2, 108.8 ± 23.7, 106.8 ± 15.7, and 101.5 ± 16.0 μm$^2$, respectively. At days 5 and 13, the surface area decreased by 6% and 26%, respectively. After day 13, the standard deviations for the cell surfaces area of the RBCs stored with CPDA-1 were 3 times larger than those for the control RBCs.

To quantitatively analyze the morphological transition from healthy discocytes to abnormal spherocytes, the sphericity was calculated from the measured cell volume and surface area (see *Materials and Methods*). A sphericity of 1 corresponds to a perfect sphere and 0 to a flat surface. At day 1, the sphericities for RBCs stored with and without CPDA-1 do not show a significant difference. However, in the absence of CDPA-1, the sphericities of RBCs exhibit large variations at day 5. After day 13, in the absence of CPDA-1, all RBCs exhibit spherocytes. The mean values of sphericity for the RBCs stored without CPDA-1 at days 1, 5, 13, 20, 27, 34, and 41 of the storage period were 0.635 ± 0.037, 0.774 ± 0.134, 0.953 ± 0.017, 0.956 ± 0.010, 0.959 ± 0.011, 0.953 ± 0.007, and 0.949 ± 0.007, respectively. In contrast, RBCs stored with CPDA-1 exhibited a prolonged transformation from discocytes to spherocytes. The mean values of sphericity for the RBCs stored with CPDA-1 at the corresponding storage periods were 0.631 ± 0.065, 0.677 ± 0.107, 0.836 ± 0.132, 0.843 ± 0.094, 0.872 ± 0.135, 0.893 ± 0.072, and 0.906 ± 0.012, respectively.

**Decreased deformability of stored RBCs**
To investigate the changes in the deformability of individual RBCs over the storage period, the dynamic membrane fluctuations of the cells were quantitatively measured. Dynamic membrane fluctuations manifest the deformability of RBC membranes, which is determined by the structures of the membrane cortex and the viscosity of the cell cytoplasm (43). The deformability of RBCs is strongly related to their ability to pass through small capillaries, and can be altered due to various pathophysiological

conditions (9, 10, 13, 23, 36-38, 45, 48).

The dynamic membrane fluctuations in RBCs were quantitatively and precisely measured using cDOT (24). The temporal full-field optical phase images of individual RBCs were measured with normal laser illumination, from which the mean and dynamic height maps of RBCs were obtained (see *Materials and Methods*). The representative mean height maps of individual RBCs at each tested storage day are presented in Fig. 3A1-A7 & C1-C7. Consistent with the results of the 3-D rendered isosurfaces (Fig. 1), the results of the mean cell shapes show that RBCs stored without CPDA-1 underwent a fast morphological transformation from discocytes to spherical shapes; RBCs lost their characteristic biconcave shapes after day 5. In contrast, RBCs stored with CPDA-1 exhibited a significantly delayed transformation; biconcave shapes were still found at day 41 for many of the RBCs.

To quantify the deformability of individual RBCs, the membrane fluctuations were calculated by spatially averaging the root-mean-squared (*RMS*) height displacements over the cell. The representative *RMS* height displacements maps of RBCs stored without and with CPDA-1 at various storage days are shown in Fig. 3B1-B7 and D1-D7, respectively. For RBCs stored without CPDA-1, the *RMS* height displacements of the spherocytes found in groups with a long storage period were significantly lower than those of the discocytes. However, the *RMS* height displacements of the RBCs stored with CDPA-1 were not dramatically decreased.

The mean membrane fluctuations of the RBCs at various days of the storage period are shown in Fig. 4. The membrane fluctuations of the RBCs stored without CPDA-1 rapidly decreased after 13 days of storage, whereas those of the RBCs stored with CPDA-1 gradually decreased and exhibited large cell-to-cell variations. The mean values for the membrane fluctuations of RBCs stored without CPDA-1 were $46.0 \pm 5.2$, $45.6 \pm 6.6$, $32.1 \pm 4.6$, $30.0 \pm 6.3$, $33.7 \pm 5.3$, $30.2 \pm 6.5$ and $30.1 \pm 4.9$ nm, and in contrast with CPDA-1 $46.8 \pm 4.1$, $47.6 \pm 7.1$, $41.0 \pm 7.1$, $42.1 \pm 7.1$, $38.2 \pm 11.0$, $39.1 \pm 9.0$ and $36.8 \pm 8.6$ nm at days 1, 5, 13, 20, 27, 34, and 41 of the storage period, respectively.

To further investigate the morphological alterations over the storage period, a correlation analysis was done between the fluctuations and sphericity. As shown in Fig. 5, the correlations between fluctuation levels and sphericity show similar trends for the RBCs with and without CPDA-1; the fluctuation level decreases as the sphericity increases. These correlations exhibited different alterations as the storage period increased for RBCs stored with or without CPDA-1 (see fitted 2-D Gaussian circles). By day 5 of the storage period without CPDA-1, the mean fluctuations of the RBCs gradually decreased as the sphericity increased from 0.6 to 1. After 2 weeks of storage, all the RBCs stored without CPDA-1 became highly spherical shapes, and their mean fluctuation levels were within 20–40 nm and significantly lower than the fluctuations of healthy discocytes at day 1. For RBCs stored with CPDA-1, the correlation clusters slowly moved toward the bottom right side in the scatter plot as the storage period increased; the clusters exhibited large deviations compared to that of the RBCs stored without CPDA-1. In addition, the results show that there exists an abrupt transition around a sphericity of 0.9 regardless of the addition of CPDA-1. For RBCs with a sphericity smaller than 0.9, the fluctuation levels increased slowly as the sphericity increased at a constant rate (approx. a 20 nm decrease in fluctuations for a 0.5 increase in sphericity). However, the fluctuation levels abruptly decreased for RBCs with a sphericity higher than 0.9. This transition in the correlation implies that the substantial changes in the deformability are associated with spherocytosis. Our results clearly show that CPDA-1 prevents spherocytosis and helps RBCs maintain a high deformability.

**Unchanged Hb content and concentration for stored RBCs**

To further investigate alternations in the cytoplasmic biochemical properties during the storage period, the hemoglobin (Hb) content and concentration in stored RBCs with or without CPDA-1 were quantified from the measured 3-D RI maps. Because the RBC cytoplasm is mainly composed of Hb and the RI of a RBC is linearly proportional to the Hb concentration, the total Hb content of individual RBCs was retrieved from the measured 2-D optical field images of the cells (2, 3, 11, 44). Then, the Hb concentrations of individual RBCs were calculated by dividing the Hb content by the cell volume (see *Materials and Methods*).

Figure 6 shows the results for the Hb contents and concentrations. After 6 weeks of storage, there were no significant changes in the Hb content for both groups of RBCs stored with and without CPDA-1 (Fig. 6A&B); the values for the Hb contents were within the reference range (28.5−33.5 pg) for healthy RBCs. The mean values for the Hb content of RBCs stored without CPDA-1 were 28.8 ± 3.6, 30.5 ± 4.7, 29.9 ± 2.9, 30.0 ± 3.4, 29.8 ± 3.7, 29.4 ± 3.6 and 31.3 ± 3.8 and in contrast with CPDA-1 28.6 ± 4.1, 29.8 ± 4.3, 28.8 ±3.3, 30.3 ± 3.8, 29.6 ± 3.2, 29.7 ± 4.0 and 30.9 ± 4.0 at days 1, 5, 13, 20, 27, 34, and 41 of the storage period, respectively. These results are consistent with a recent work using digital holographic microscopy (33).

The Hb concentrations for both groups of RBCs also did not significantly change over 41 days of storage (Fig. 6B&D). This can be easily understood due to the insignificant changes in both the Hb contents and cellular volumes. The mean values for the Hb concentrations of RBCs stored without CPDA-1 were 32.2 ± 1.9, 34.9 ± 4.0, 37.1 ± 1.9, 36.2 ± 2.2, 36.4 ± 2.4, 34.3 ± 1.9 and 35.4 ± 1.8 pg/dL, and with CPDA-1 were 34.8 ± 2.5, 34.6 ± 3.2, 34.5 ± 3.0, 35.4 ± 2.5, 34.7 ± 3.4, 35.7 ± 2.7 and 36.7 ± 4.4 pg/dL, at days 1, 5, 13, 20, 27, 34, and 41 of the storage period, respectively. All the mean values for the Hb concentrations in the two groups were within the reference range for healthy RBCs (33–36 g/dL). This result suggests that the cytoplasmic Hb proteins were not significantly degraded or removed during the 6 weeks of storage, despite the dramatic alterations in the morphological and mechanical properties. The unchanged Hb contents and concentrations in the RBCs stored with CPDA-1 were consistent with previous CBC measurements (1).

**Conclusion & Discussion**

In this paper, we investigated alterations in individual human RBCs during blood storage. With 3-D QPI techniques, the morphological (cellular volume, surface area, and sphericity), biochemical (Hb content, and Hb concentration) and mechanical properties (membrane fluctuation) of individual RBCs stored with and without CPDA-1 were systemically quantified.

Our optical measurements showed no significant changes in cellular volumes, Hb contents and Hb concentrations during 6 weeks of storage for RBCs stored without CPDA-1. However, the surface areas of the RBCs significantly decreased during the first two weeks, showing sphericity values of unity. The morphologies of the RBCs dramatically transformed in less than 1 week of storage. In the absence of CPDA-1, 60% of the RBCs stored without CPDA-1 were a non-discocyte shape at day 5; after day 13, all the RBCs became spherical shapes. These results – the decrease in cell membrane area and the increase in sphericity – imply that spherocytosis is induced by vesiculation in the stored RBCs. This storage-induced spherocytosis seems to cause a significant decrease in cell deformability; membrane fluctuations in the RBCs decreased by 53% after two weeks of storage.

In the RBCs stored in the presence of CPDA-1, the morphological transformation to spherocytosis was significantly delayed compared to the RBCs stored without CPDA-1. In the presence of CPDA-1, 80% of the RBCs remained as discocytes at day 5. Furthermore, there was a large fraction of discocytes after 6 weeks of storage (38% of the RBCs had a sphericity less than 0.9), although the number of sphere-like RBCs gradually increased as the storage period increased. This decreasing trend in cell deformability is qualitatively consistent with a previous report (6), in which the averaged values of the deformation index of the RBCs were significantly reduced after two weeks of storage (approx. 39%).

It is well known that CPDA-1 extends the survival days of stored RBCs by providing adenine to maintain cytoplasmic ATP levels (17, 49). Our results clearly show that distinct trends in stored RBCs - morphological transformation from discocytes to spherocytes as well as a decrease in membrane deformability – were considerably reduced in the presence of CPDA-1 at the individual cell level. This result agree with a previous work using a narrow micro channel mimicking microcirculation (27). In particular, the decreased amplitudes in dynamic membrane fluctuation decreases over storage time (Figs. 4A-B) are consistent with those of a previous work in which cytoplasmic ATP was depleted chemically and metabolically (4, 7, 36, 47). More importantly, our results suggest the presence of sub-groups in RBCs exhibiting various responses to CPDA-1. For example, the sphericities and dynamic membrane fluctuations after two weeks of storage showed highly dispersed values ranging from healthy levels to severe spherocytes. This large cell-to-cell variations in the responses to CPDA-1 could be from various factors, including the different ages of the individual RBCs or various cellular levels of potassium and 2,3-DPG in RBCs (5).

Our results suggests measurements of sphericity can be exploited to address the expected duration of RBCs in clinic. Traditional CBC measurements do not provide information about sphericity or cell surface area. Hence, it remained challenging to expect the survival duration of RBCs in stored blood.

Currently, reticulocyte count can be used for addressing the fraction of young RBCs. However, the fraction of old RBCs in blood cannot be measured. In the view of clinical hematology, the results presented in this work suggests that sphericity or surface area is strongly correlated with the viability of RBCs and can also be used as a proxy for the cell age of RBCs. In other words, measurements of cell sphericity may enable to provide information about the expected survival duration of RBCs. In particular, the sphericity higher than 0.9 or the membrane fluctuations of 35–40 nm or below could be used for criteria to identity inviable RBCs.

In this paper, we presented the optical measurements for various parameters of stored RBCs and investigated the progressive alterations of these retrieved parameters. We believe the present method will be useful in clinical applications for hematology. For example, the present method can be used to non-invasively assess blood preservatives or the functionalities of stored blood. Using the recently developed quantitative phase imaging units (21, 30) and a simple illumination scheme (50), existing optical microscopes in a laboratory can be converted into quantitative phase microscopy, which can possibly extend the applicability of the present method. Furthermore, the optical measurements of RBC parameters presented in this paper can be combined with flow cytometry techniques and fast 3-D tomographic measurements (20) such that individual RBCs can be sorted according to their parameters.

**Materials and Methods**

**Ethics statements and Blood sample preparation** Human blood studies were conducted according to the principles of the Declaration of Helsinki and were approved by the responsible ethics committee of Eulji University Hospital (IRB project number: 2012-0128, Daejeon, Republic of Korea). Blood samples from a healthy adult was obtained via regular course of patient care after approval in accordance with the procedures of IRB for the remaining blood. Blood was collected into EDTA treated anticoagulant tube and divided into two tube. In one of tubes, CPDA-1(citrate-phosphate-dextrose with adenine, C4431, Sigma-Aldrich, U.S.A.) was added into whole blood with 1:9 v/v ratio. The other tube without CPDA-1 was kept for control sample. Both two tubes were stored at 4°C and agitated occasionally. Every few days up to 41 days, the blood from each tubes were tested for quality check using 3-D optical tomography. All experimental protocol were proved by the institutional review board of KAIST (KH2013-22). For optical measurement, blood was further diluted 300 times in Dulbecco's PBS buffer (Gibco®, New York, U.S.A.).

**Common-path diffraction optical tomography (cDOT)** The Common-path diffraction optical tomography is a 3-D QPI technique that can measure the 3-D distribution of RI of individual cells (24, 51). As an illumination source, a diode-pumped solid state laser ($\lambda$ = 532 nm, 50 mW, Cobolt, Solna, Sweden) was used. The sample, diluted blood sandwiched between two cover glasses with 25×50 mm (C025501, MATSUNAMI GLASS Ind., LTD., JAPAN), is placed between the condenser lens (UPLSAPO 60×, numerical aperture (N.A.) = 0.9, Olympus, Japan) and objective lens (UPLSAPO 60×, N.A. = 1.42, Olympus, Japan).

For 3-D RI tomography, the optical fields obtained with various incident illumination angles were measured by employing the common-path laser-interferometric microscopy (22, 24, 51). By rotating a two-axis galvanometric mirror (GVS012/M, Thorlabs, USA), incident angles of illumination beams to a sample is controlled. The second galvanometer mirror reflected the beam from a sample to have the same optical path regardless of incident illumination angles.

Holograms of the sample were recorded, employing the principle of common-path interferometry (39, 42). After the second galvanometer mirror, a diffraction grating (92 grooves mm$^{-1}$, #46-072, Edmund Optics Inc., NJ, U.S.A.) spatially split the scattering beams and then, spatially filtered 0$^{th}$ order beam as a reference was interfered with the 1$^{st}$ order beam as a sample beam. Then, the interferograms were recorded on high-speed sCMOS camera (Neo sCMOS, ANDOR Inc., Northern Ireland, UK) while the incident beam was scanning spirally with 300 different angles. The total magnification was 240 by an additional 4-$f$ system. Then, optical field images, containing both the amplitude and phase maps, of the sample were retrieved using a phase retrieval algorithm (12). From the retrieved optical fields, 3-D RI distribution of sample was reconstructed using optical diffraction tomography algorithm. Detailed information about the reconstruction algorithm can be found elsewhere (22, 31).

**Analysis of the red cell parameters** From the measured 3D RI tomograms and the 2D dynamic membrane fluctuations, six red cell parameters are retrieved, including morphological (cell volume, surface area and sphericity), chemical (Hb content and Hb concentration), and mechanical (membrane fluctuation) parameters (25, 34, 35, 56). To measure the morphological parameters, we used the reconstructed 3-D RI maps (25). The whole volume of a RBC was calculated by integrating all voxels inside individual RBCs. The space corresponding to the cytoplasm of a RBC was selected by RI with a higher value than the threshold. The threshold was defined by 50% of RI difference between the maximum RI of the cell $n_{cell\_max}$ and surrounding medium $n_m$ for determine the cell boundary, i.e. $n_{thresh} = n_m + 0.5 \cdot (n_{cell\_max} - n_m)$. Then, the total number of voxels was multiplied by the magnification of the optical system to translate in a length scale. Next, for the surface area measurements, the isosurfaces of individual RBCs were reconstructed from the volume data of the 3-D RI maps. The surface area of the isosurface was measured with the sum of the areas of all the patch faces, which were broken down into small triangular pieces. In addition, the sphericity $SI$, a dimensionless quantity ranging from 0 to 1, was obtained as follows: $SI = \pi^{1/3}(6V)^{2/3}/A$ where $V$ is the volume and $A$ is the surface area (25, 52, 55).

To obtain Hb content of individual RBCs, the measured 2-D phase at the normal angle was used. The Hb content of a RBC was obtained from integrating the 2-D optical phase over the entire cell area and with the RI increased of the Hb proteins, given as follows: Hb content = $(\lambda/2\pi\alpha)\int\Delta\phi(x,y)dA$, where $\lambda$ is the wavelength of the illumination laser light (532 nm), $\alpha$ is the RI increment (0.2 mL/g) (11, 44), and $\Delta\phi(x,y)$ is 2-D optical phase. In addition, the Hb concentration in a RBC was obtained from the Hb content divided by the cellular volume (19, 40).

The dynamic membrane fluctuations in RBCs can be quantitatively and precisely measured using cDOT (24, 43). Dynamic full-field optical phase images of a RBC $\Delta\phi(x, y, t)$ can be measured with a normal-angle laser illumination, from which dynamic height maps of the RBC can be calculated as, $h(x, y, t) = [\lambda/(2\pi\cdot\Delta n)]\Delta\phi(x, y, t)$, and $\Delta n = \langle n(x, y, z)\rangle - n_m$ is a difference between the mean RI of RBC cytoplasm $\langle n(x, y, z)\rangle$ and surrounding buffer medium $n_m$.

The representative cell height images of individual cells in each groups are calculated as the temporally averaged cell heights, $h_m(x, y) = \langle h(x, y, t)\rangle$, and are presented in Figs. 3A & C. To measure the mechanical parameter, we calculated the dynamic membrane fluctuation from the successively measured instantaneous height map $h(x,y;t)$, given as, $h(x,y;t) = [\lambda/(2\pi\cdot\Delta n)]\Delta\phi(x,y;t)$. The values for the membrane fluctuation were calculated by averaging the root-mean-square of the height displacement over the cell area, given as: $\Delta h(x,y) = \langle[h(x,y;t) - h_m(x,y)]^2\rangle^{1/2}$, where $h_m$ is the time averaged height at the cell surface.

**ACKNOWLEDGEMENTS**
This work was supported by KAIST-Khalifa University Project, APCTP, the Korean Ministry of Education, Science and Technology, and the National Research Foundation (2012R1A1A1009082, 2014K1A3A1A09063027, 2013M3C1A3063046, 2012-M3C1A1-048860, 2014M3C1A3052537).

**AUTHOR CONTRIBUTIONS**
H.P., S.J. and Y.P. developed the experimental idea. H.P., Y.L. and K.K. performed the experiments and analyzed the data. M.J., Y.S. and S.J. prepared blood samples. Y.S., S.J., and Y.P. supervised the study. All authors discussed the experimental results and wrote the manuscript.

**COMPETING FINALCIAL INTERESTS**
The authors declare no competing financial interests.

References

**Figures with legends**

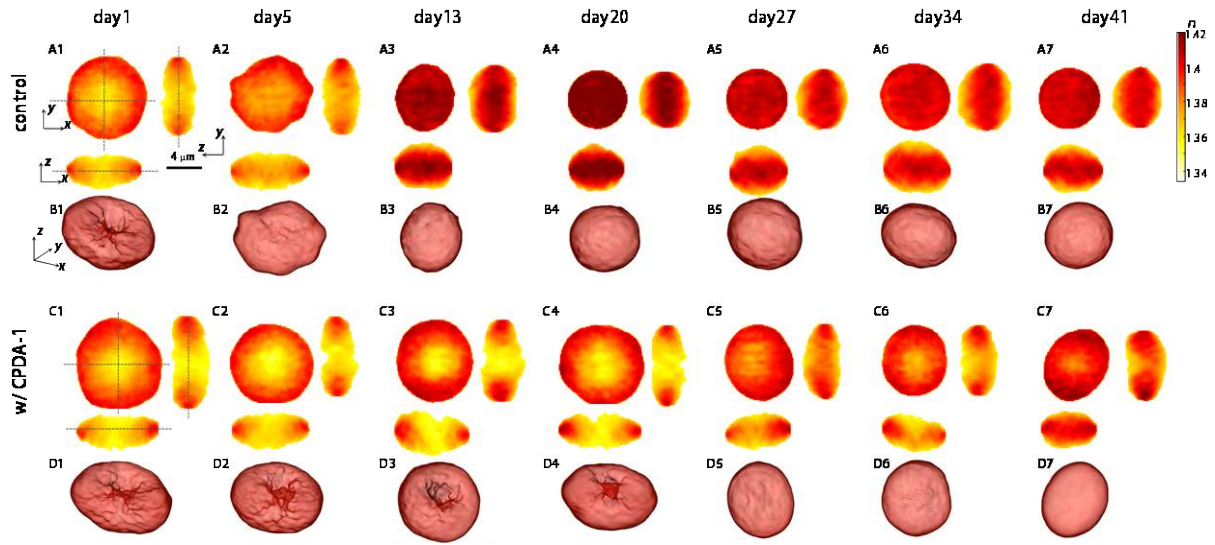

**Fig. 1** Cross-sectional slices of reconstructed RI tomograms of stored RBCs without CPDA-1 as the control (A1-A7) and with CPDA-1 (C1-C7) at days 1, 5, 13, 20, 27, 34, and 41 of the storage period, respectively, in the *x-y* (left), *y-z* (right), and *x-z* (below) plane (B1-B7 and D1-D7). Corresponding rendered isosurfaces of 3-D RI maps for the control and CPDA-1, respectively (n > 1.360).

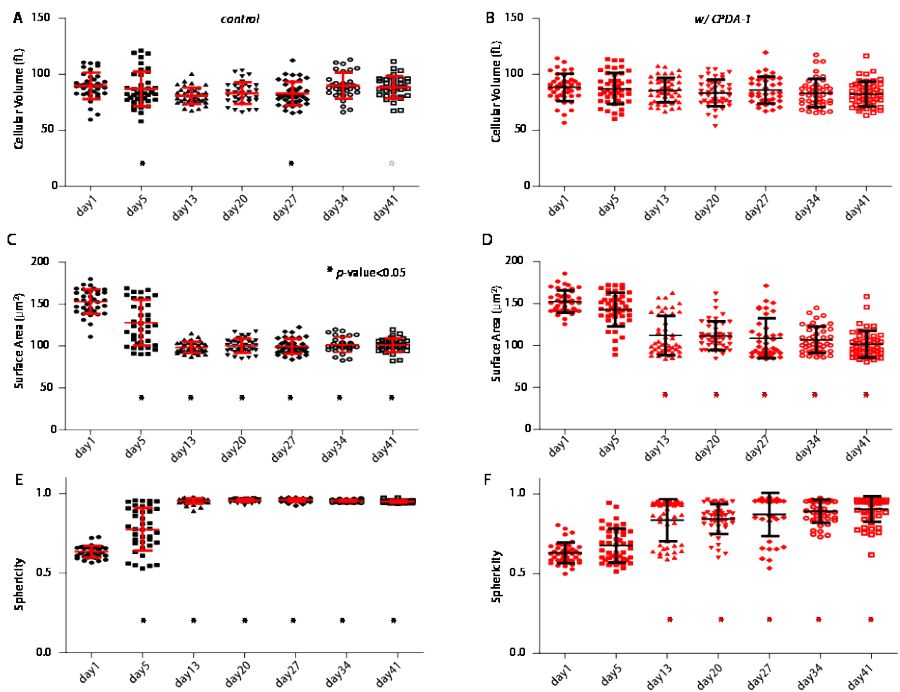

**Fig. 2** Morphological red cell indices of the stored RBCs without CPDA-1 as the control (A, C and E) and with CPDA-1 (B, D, and F) at days 1, 5, 13, 20, 27, 34 and 41 of the storage period. (A&B) Cell volumes. (C&D) Cell surface areas. (E&F) sphericities. Each symbol corresponds to an individual RBC measurement. The horizontal solid line is the mean value with an error bar showing the standard deviation.

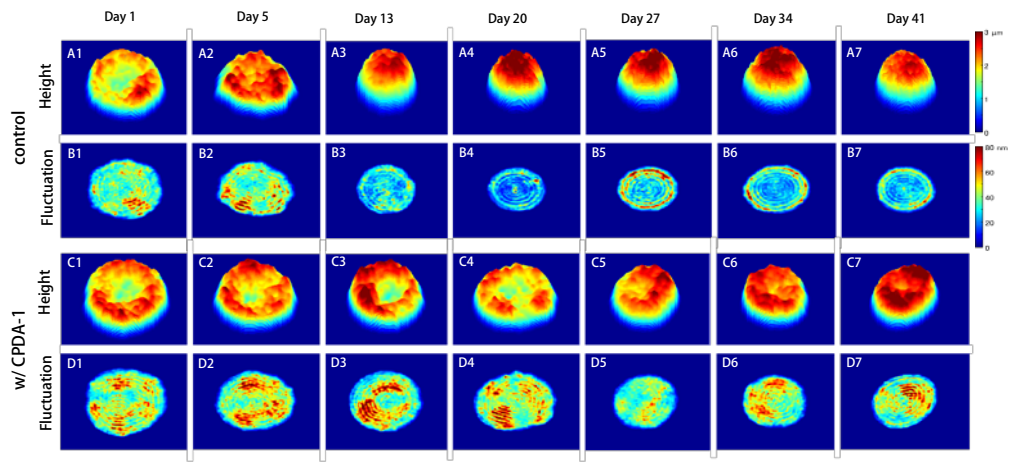

**Fig. 3** Representative 2-D topographical images of stored RBCs without CPDA-1 as the control (A1-A7) and with CPDA-1 (C1-C7) at days 1, 5, 13, 20, 27, 34, and 41 of the storage period (left to right), respectively. (B & D) Corresponding membrane fluctuation maps.

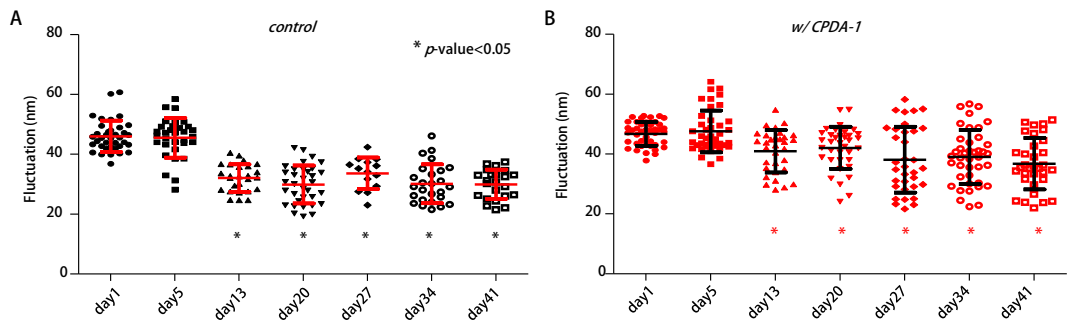

**Fig. 4** Mean membrane fluctuations of individual stored RBCs without CPDA-1 as the control (A) and with CPDA-1 at days 1, 5, 13, 20, 27, 34, and 41 of the storage period.

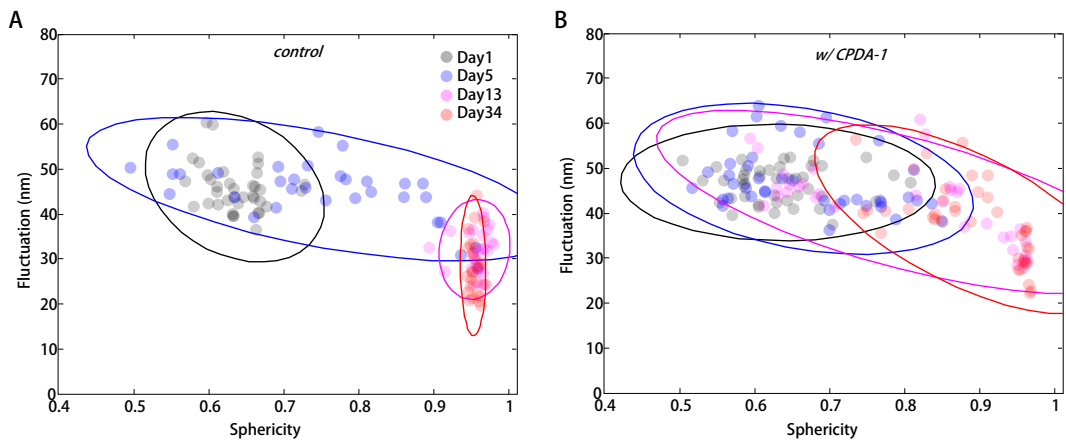

**Fig. 5** Correlation maps between sphericity and mean membrane fluctuations of (A) stored individual RBCs without CPDA-1 and (B) with CPDA-1 at days 1, 5, 13, 20, 27, 34, and 41 of the storage period.

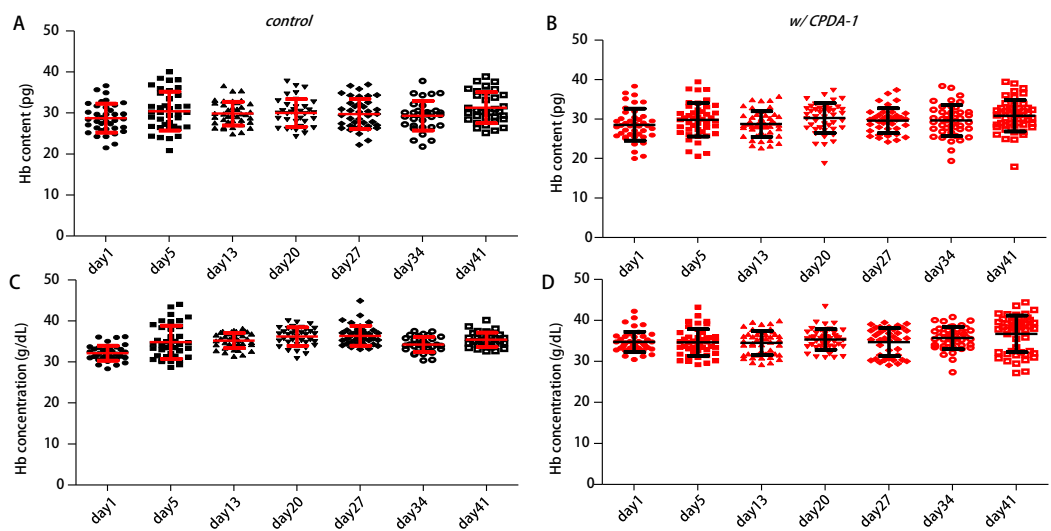

**Fig. 6** Hb content and concentration in the stored RBCs without CPDA-1 as the control (A and C) and with CPDA-1 (B and D), respectively, at days 1, 5, 13, 20, 27, 34, and 41 of the storage period.